# 3plex Web: An Interactive Platform for RNA:DNA Triplex Prediction and Analysis

Marco Masera[1,†], Chiara Cicconetti[1,†], Francesca Ferrero, Salvatore Oliviero[1] and Ivan Molineris[1,*]

[1]Dipartimento di Scienze della Vita e Biologia dei Sistemi and MBC, Università di Torino, Via Accademia Albertina 13, 10126 Torino, Italy.

*To whom correspondence should be addressed.

†These authors contributed equally to this work.

## Abstract

**Summary:** Long non-coding RNAs (lncRNAs) exert their functions by cooperating with other molecules including proteins and DNA. Triplexes, formed through the interaction between a single-stranded RNA (ssRNA) and a double-stranded DNA (dsDNA), have been consistently described as a mechanism that allows lncRNAs to target specific genomic sequences *in vivo*. Building on the computational tool 3plex, we developed 3plex Web, an accessible platform that enhances RNA:DNA triplex prediction by integrating interactive visualization, statistical evaluation, and user-friendly downstream analysis workflows. 3plex Web implements new features such as input randomization for statistical assessments, interactive profile plotting for triplex stability, and customizable DNA Binding Domain (DBD) selection. This platform enables rapid analysis through PATO, substantially reducing processing times compared to previous methods, while offering Snakemake workflows to integrate gene expression data and explore lncRNA regulatory mechanisms.

**Availability and implementation:** 3plex Web is freely available at https://3plex.unito.it as an online web service. The source code for 3plex is available at https://github.com/molinerisLab/3plex, paired with a definition file to set up the application into a Singularity image.

**Contact:** ivan.molineris@unito.it

## 1. Introduction

Triplexes are molecular complexes resulting from the interaction of ssRNA and dsDNA molecules. The formation of these structures requires the Watson-Crick interacting nucleotides of the DNA to establish an additional hydrogen bond with the nucleotides of the RNA (Hoogsteen pairing) [1]. Triplexes principally occur following a canonical set of rules: a purine-rich portion of the DNA can bind a pyrimidinic or mixed portion of the RNA in a parallel direction [C:G-C, U:A-T, G:G-C] and, alternatively, a purinic or mixed portion of the RNA in an anti-parallel direction [G:G-C, A:A-T, U:A-T] [2]. The triplex-forming subsequence on the ssRNA is denoted as Triplex Forming Oligonucleotide (TFO) and its cognate substring on the dsDNA is named Triplex Target Site (TTS). Taken together, the TSS and the associated TFO combine to form a triplex (TPX).

The interest in these peculiar structures began with the necessity to create biotechnological instruments that recognize DNA sequences with high specificity for therapeutic purposes [3]. LncRNAs emerged as important players in the regulation of gene expression and in the nuclear structural organization [4]. Mechanistically, lncRNAs exert their functions by cooperating with other molecules, and ssRNA:dsDNA TPX formation has been consistently described as a mechanism that allows lncRNAs to target specific genomic sequences *in vivo* [5, 6, 7]. Given the increasing evidence of the biological importance of these interactions, we previously developed 3plex, a tool for the *in-silico* prediction of TPXs [8]. With 3plex, we enhanced the state-of-the-art sequence-based TPX prediction algorithm (Triplexator [9]) with relevant biophysical knowledge of these structures. 3plex differentiates itself from other methods by enriching sequence-based prediction with data coming from denaturation experiments, fundamental for determining the thermal stability of the structures [10], [11]. The importance of this measurement is demonstrated by recent research that has expanded its understanding [12].

Moreover, we integrated the prediction of the RNA secondary structure [13].

In this paper we present 3plex Web, a refined and enhanced version of our tool, designed to elevate the precision, efficiency, and accessibility of RNA:DNA TPX prediction and analysis. The major improvements are:

- Achieves 100-fold increase in speed by replacing Triplexator with PATO [14] a reimplementation of Triplexator that significantly reduces runtime, enabling the use of computationally

- expensive yet more accurate prediction parameters we already elucidated in our previous work [8].
- Enables the execution of analyses on the web without the need for complex software installations or advanced computational skills, thereby making TPX analysis accessible to all biologists.
- Provide interactive visualization tools for result interpretation.
- Introduce a randomization procedure to statistically evaluate the biological significance of predicted TPXs.
- Include a suite of predefined Snakemake workflows for downstream analyses, similarly to TDF [15] approach but also considering thermal stability. This allows a seamless integration of "omics" datasets, such as RNA-seq and ChIRP-seq, to investigate the regulatory potential of lncRNAs. These workflows are accessible via the command-line interface, with selected features integrated into the web platform for enhanced usability.

In our previous work we extensively demonstrated that 3plex accuracy outperformed the other existing tools (including Triplexator and TriplexAligner [16]). Compared to 3plex Web, TRIPBASE [17] provides a precomputed database of TPX predictions across the human genome, it offers remote execution of the Triplexator (slower and less accurate compared to 3plex) on user-provided sequences, shows accessibility problems and limited interactive exploration capabilities. Triplex Domain Finder [15] requires complex local installation of software and reference genomic data, it is based on Triplexator, does not include thermal stability or secondary structure considerations, and has limited interactive exploration capabilities. Genna et al recently developed a new TPX predictor based on machine learning approaches [12], but they did not publish the software nor the model parameters, making challenging to compare or to use it in other contexts.

## 2. 3plex Web

To submit a 3plex Web prediction job, users must provide: 1) one ssRNA sequence in FASTA format and, 2) a set of dsDNA sequences in multi-FASTA format. For human and mouse data, the ssRNA input can be selected from the searchable list of GENCODE-annotated transcripts, while the dsDNA input can be provided in BED format or as a list of genes. Users can adjust key parameters influencing TPX prediction, such as the minimum TPX length and masking of the ssRNA based on the secondary structure prediction. Optionally, a randomization procedure can generate a randomized set of dsDNA sequences, enabling a robust statistical comparison between observed TTS profiles and expected distributions.

To evaluate the propensity of ssRNA regions to form biologically relevant TPXs, 3plex Web provides useful visualization functionalities (Fig. 1). The "TTS count" plot displays the total number of TTSs identified along the ssRNA sequence together with the thermal stability of the best predicted TPX. If the randomization procedure is enabled, the plot also displays the expected TTS count in randomized sequences and reports empirical p-values. Although highly stable TPXs typically require the involvement of long stretches of the ssRNA, we previously demonstrated that short and less stable TPXs can play regulatory roles in gene expression, like transcription factor binding events [5], [18][19]. To accommodate these considerations, 3plex Web allows users to dynamically adjust the stability threshold in real-time, facilitating an exploratory analysis of TPX formation potential across the entire ssRNA sequence.

Since TFOs are expected to be formed by unfolded portions of the ssRNA, the "Secondary structure" plot displays the propensity of the sequence to be double stranded in the RNA secondary structure [13]. If human or mouse species is selected, two additional plots visualize the sequence phylogenetic conservation score [20] and the repetitive element position.

To assist in the identification of functional TPX-forming regions, 3plex Web enables the interactive detection of DNA Binding Domains (DBDs)—ssRNA regions predicted to engage in sequence-specific TPXs. In contrast with other tools like Triplex Domain Finder [15], which automatically define the DBDs, 3plex Web provides users with the flexibility to manually adjust DBD boundaries, refining their position and width for more tailored analyses. For each DBD, users can view and download the complete list of TFO-TTS interactions in tabular format. For human and mouse data, users can access a searchable table detailing key properties for each dsDNA region, including genomic coordinates, associated genes, best stability score, best PATO score and normalized thermal stability. Selecting a specific dsDNA region dynamically updates the profile plots, restricting the analysis to TPXs associated with the chosen genomic region. Additionally, a direct link to the UCSC Genome Browser allows users to inspect the predicted TPX binding sites within their genomic context. A full overview of the 3plex Web interface is provided in Supplementary Material S1.

## 3. 3plex Snakemake workflows

Investigating the mechanism of action of lncRNAs requires assessing the biological significance of their predicted interactions with target molecules to formulate meaningful hypotheses [21]. To support this, we have implemented downstream analysis pipelines that provide insights into the TPX-mediated functionality of an ssRNA. These pipelines enable the integration of omics data, such as RNA-seq and ChIRP-seq, which are commonly employed to explore lncRNA function.

The random region test determines whether a specific set of genomic regions (e.g., ChIRP-seq peaks) could arise from a TPX-mediated interaction with a specific lncRNA subregion i.e., a DBD. To assess this, $N$ randomized versions of the target regions are generated using *bedtools shuffle* [22], excluding ENCODE blacklists and genomics gaps. 3plex is then executed on both the original target regions and the randomized regions, and the TTS count and thermal stability are computed for each nucleotide of the ssRNA. DBDs are identified based on the overlapping TFOs found in the target regions. For each DBD, the upper quartile of the TTS count and the thermal stability of the included nucleotides are extracted. These values are then compared with those derived from the randomized control regions to compute empirical p-values, providing statistical significance for TPX-mediated interactions.

The promoter TPX stability test evaluates the potential role of a candidate ssRNA in regulating a specific set of genes, such as those identified as differentially expressed in knock-out experiments. By integrating gene expression data, this workflow assesses whether TPXs are enriched at promoters of regulated genes. Users provide a background gene set (e.g., all the expressed genes in the system), a list of genes of interest (e.g., differentially expressed genes following ssRNA knockout), and the ssRNA sequence. Promoters are pre-defined as spanning -1500 to +500 bp relative to transcription start sites, based on Matched Annotation from NCBI and EMBL-EBI (MANE) [23] 3plex predicts TPX formation with these promoters and evaluates the stability of interactions between the ssRNA and promoters of genes of interest compared to all other genes using a Mann-Whitney test. Additionally, a Gene Set Enrichment Analysis [24], [25] ranks genes by their TPX stability score, assessing the enrichment significance of high or low stability interactions at the promoters of selected genes. The leading-edge table provides a selection of candidate target genes for further investigation.

Both pipelines are designed for command-line execution using Snakemake. Additionally, the promoter TPX stability test is integrated into 3plex Web for remote job execution.

## 4. Implementation

3plex Web is a platform that provides a graphical user interface for the 3plex core application on a remote cluster, allowing users to interactively navigate the results. It consists of four distinct components: a web client based on the Angular framework, a public (frontend) server based on the Django rest framework, a private (backend) server and the core 3plex application. The web client offers a graphic interface for the user and is the entry point for the application. The frontend server offers public APIs to the client, manages user data, handles data visualization logic, caches result and securely stores users' job information. The backend is a stateless server deployed on an HPC cluster; it manages the enqueueing of the 3plex jobs on the cluster using SLURM and holds back the results when ready. The jobs are then executed by the 3plex core application, enclosed in a singularity container, on the first available node.

Upon job submission, the platform generates a unique token, allowing users to retrieve their results when available. Alternatively, users can provide an email address to receive a notification once the job is complete. Job data are retained on the server for up to four weeks. The platform offers a functionality to export the job data, allowing to restore an expired job.

The 3plex core application, implemented in Snakemake [26] has been optimized by integrating PATO [14], improving computational efficiency compared to the previous Triplexator-based implementation. PATO maintains the canonical set of rules for TPX identification while significantly reducing computation time and delivering easily interpretable output.

3plex introduces an additional data structure, the profile, which summarizes TTS counts across all stability thresholds for each segment of the ssRNA sequence. These profiles are stored as sparse matrices compressed in msgpack format, enabling direct access via the web client. Moreover, 3plex now supports randomizations of dsDNA sequences, generating a second profile showing the average, the variance and the quartiles of expected TPX count over the randomized DNAs.

Command-line functionalities leverage Snakemake workflow manager and are distributed via GitHub. This modular tool seamlessly integrates into biological data analysis pipelines, enabling TPX-focused downstream analyses.

## Funding

This work has been supported in part by grant 2024-342822 (5022) GB-1609971 from the Chan Zuckerberg Initiative DAF, an advised fund of Silicon Valley Community Foundation.

*Conflict of Interest:* none declared.